\documentclass[a4paper,12pt]{article}
\usepackage[latin1]{inputenc}
\usepackage[dvips]{graphicx}

\newcommand{\cent}{\centerline}
\newcommand{\disp}{\displaystyle}
\newcommand{\drm}{{\rm d}}
\newcommand{\erm}{{\rm e}}

\newcommand{\noi}{\noindent}

\newcommand{\Ocal}{{\rm {\cal O}}}
\newcommand{\Acal}{{\rm {\cal A}}}
\newcommand{\Lcal}{{\cal L}}
\newcommand{\al}{\alpha}
\newcommand{\inr}{{\rm in}}
\newcommand{\fir}{{\rm fin}}

\begin{document}

\hfill{Preprint NSF-ITP-02-62}

\

\

\

\begin{center}
\cent{{\bf A SIMPLE QUANTUM EQUATION }}

\cent{{\bf FOR DECOHERENCE AND DISSIPATION}$^{\: (\dag)}$}
\footnotetext{$^{\: (\dag)}$ This reasearch was supported in part by the N.S.F.
under Grant No.PHY99-07949; and by INFN and Murst/Miur (Italy). \ E-mail for contacts:  recami@mi.infn.it }

\

\end{center}

\vspace*{5mm}

\cent{Erasmo Recami}

\centerline{{\em Facolt\`{a} di Ingegneria, Universit\`{a} statale di Bergamo,
Dalmine (BG), Italy;}} \centerline{{\em INFN---Sezione di Milano, Milan,
Italy; \ {\rm and}}} \centerline{{\em Kavli Institute for Theoretical Physics, UCSB,
CA 93106, USA.}}

\

\cent{Ruy H. A. Farias}

\cent{{\em LNLS - Synchrotron Light National Laboratory;
Campinas, S.P.; Brazil.}}
\cent{ruy@lnls.br}

\

{\em Abstract --} \
Within the density matrix formalism, it is shown that a simple way to get
decoherence is through the introduction of a ``quantum" of time
({\em chronon\/}): which implies replacing the differential
Liouville--von~Neumann
equation with a finite-difference version of it.  In this way, one is given
the possibility of using a rather simple quantum equation to describe the
decoherence effects due to dissipation.
Namely, the mere introduction (not of a ``time-lattice", but simply)
of a ``chronon" allows us to go on from differential to finite-difference
equations; and in particular to write down the quantum-theoretical equations
(Schroedinger equation, Liouville--von~Neumann equation,...) in three different
ways:
``retarded", ``symmetrical", and ``advanced". One of such three formulations
---the {\em retarded} one--- describes in an elementary way a system which
is exchanging (and losing) energy with the environment; \ and in its
density-matrix
version, indeed, it can be easily shown that all non-diagonal terms go to
zero very rapidly.

\

{\em keywords:}
quantum decoherence, interaction with the environment, dissipation, quantum
measurement
theory, finite-difference equations, chronon, Caldirola, density-matrix
formalism, Liouville--von~Neumann equation

\vfill\newpage

\section*{Introduction}

In this note, we briefly mention the consequences of the introduction of a
quantum of time $\tau_0$
in the formalism of non-relativistic quantum mechanics, by referring
ourselves, in particular, to the theory of the {\em chronon} as proposed
by P.Caldirola. Let us recall that such an interesting ``finite difference" theory, forwards
---at the classical level--- a solution for the motion of a particle endowed
with a non-negligible charge in an external electromagnetic field, overcoming
all the known difficulties met by Abraham--Lorentz's and Dirac's approaches
(and even allowing a clear answer to the question whether a free falling
charged particle does or does not emit radiation), and ---at the quantum
level--- yields a remarkable mass spectrum for leptons.

In unpublished work [cf. the e-print quant-ph/9706059, and the subsequent
Report IC/98/74 (I.C.T.P.; Trieste, 1998), where also extensive references
can be found], after having reviewed Caldirola's
approach, we\cite{RUY1} worked out, discussed, and compared to one another the {\em new}
representations of Quantum Mechanics (QM) resulting from it, in the
Schr\"odinger, Heisenberg and density--operator (Liouville--von~Neumann)
pictures, respectively.

Let us stress that, for each representation, three ({\em retarded, symmetric} and
{\em advanced}) {\em formulations} are possible, which refer either to
times $t$ and $t-\tau_0$, or to times $t-\tau_0/2$ and $t+\tau_0/2$, or to
times $t$ and $t+\tau_0$, respectively. \ It is interesting to notice that,
when the chronon tends to zero, the ordinary QM is obtained as the limiting
case of the ``symmetric" formulation only; while the ``retarded" one does
naturally appear to describe QM with friction, i.e., to describe
{\em dissipative} quantum systems (like a particle moving in an absorbing
medium). In this sense, {\em discretized} QM is much richer than the
ordinary one.

In the mentioned unpublished work\cite{RUY1}, we have also obtained the
(retarded) finite--difference Schr\"odinger equation within the Feynman
path integral approach, and studied some of its relevant solutions.  We have
then derived the time--evolution operators of this discrete
theory, and used them to get the finite--difference Heisenberg
equations.\footnote{When discussing therein the mutual compatibility of
the various pictures listed above, we found that they {\em can} actually be
written in a form such that they result to be equivalent (as it happens in
the ``continuous" case of ordinary QM), even if the Heisenberg
picture cannot be derived by ``discretizing" directly the ordinary
Heisenberg representation.} \ Afterwards, we have studied some typical
applications and examples: as the free particle, the harmonic oscillator and
the hydrogen atom; and various cases have been pointed out, for which the
predictions of discrete QM differ from those expected from ``continuous" QM.

We want here to pay attention to the fact that, when applying the density
matrix formalism to the solution of the {\em measurement problem} in QM,
very interesting results are met, as, for instance, a natural explication
of the ``decoherence"\cite{DECO2} due to dissipation: which seem to reveal
the power of dicretized (in
particular, {\em retarded\/}) QM.

\section{Outline of the classical approach}

If $\rho$ is the charge density of a particle on which an external
electromahnertic field acts, the famous Lorentz's force law

\begin{equation}
\bf{f} = \rho \left({\bf{E} + \frac{1}{c}\bf{v}\wedge\bf{B}}\right) \ ,
\end{equation}  

\noindent
is valid only when the particle charge $q$ is negligible with respect to the
external field sources.  Otherwise, the classical problem of the motion of a
(non-negligible) charge in an electromegnetic field is still an open question.
For instance, after the known attempts by Abraham and Lorentz, in 1938
Dirac\cite{DIRA3} obtained and proposed his famous classical equation

\begin{equation}
m \frac{\drm u_{\mu}}{\drm s} = F_{\mu} + \Gamma_{\mu} \ ,
\label{eq:dirrel}
\end{equation}  

\noi
where $\Gamma_{\mu}$ is the Abraham 4-vector

\begin{equation}
\Gamma_{\mu}=\frac{2}{3}\frac{e^2}{c}\left({\frac{\drm^2 u_{\mu}}{\drm s^2}
+\frac{u_{\mu}u^{\nu}}{c^2} \frac{\drm ^2u_{\nu}}{\drm s^2}}\right) \ ,
\end{equation}  

\noi
that is, the (Abraham) reaction force acting on the electron itself;  and
$F_{\mu}$ is the 4-vector that represents the external field acting on
the particle

\begin{equation}
F_{\mu}=\frac{e}{c} F_{\mu \nu} u^{\nu} \ .
\end{equation}  

\noindent
At the non-relativistic limit, Dirac's equation goes formally into
the one previously obtained by Abraham--Lorentz:

\begin{equation}
m_0\frac{\drm{\bf v}}{\drm t}-\frac{2}{3}\frac{e^2}{c^3}\frac{\drm^2{\bf v}}
{\drm t^2}=
e\left({{\bf E}+\frac{1}{c} {\bf v}\wedge {\bf B}}\right) \ .
\label{eq:dirnrel}
\end{equation}     

\noindent
The last equation shows that the reaction force equals \ ${2 \over 3} \;
{e^2 \over c^3} \; {\drm^2 {\bf v} \over {\drm} t^2}$.

Dirac's dynamical equation (2) presents, however, many troubles, related to
the infinite many non-physical solutions that it possesses. \ Actually, it is
a third--order differential equation, requiring three initial conditions for
singling out one of its solutions. \ In the description of a {\em free} electron,
e.g., it yields ``self-accelerating" solutions ({\em runaway
solutions\/}), for which velocity and acceleration increase
spontaneously and indefinitely. Moreover, for an electron submitted to an
electromagnetic pulse, further non-physical solutions appear, related this
time to {\em pre-accelerations}: If the electron comes from
infinity with a uniform velocity $v_0$ and, at a certain instant of time
$t_0$, is submitted to an electromagnetic pulse, then it starts accelerating
{\em before} $t_0$. \ Drawbacks like these motivated further attempts to find
out a coherent (not pointlike) model for the classical electron.

Considering elementary particles as points is probably the sin plaguing
modern physics (a plague that, unsolved in classical physics, was transferred
to quantum physics).  One of the simplest way for associating a discreteness
with elementary particles (let us consider, e.g., the electron) is just via
the introduction (not of a ``time-lattice", but merely) of a ``quantum" of
time, the {\em chronon}, following Caldirola.\cite{CALREV4} \
Like Dirac's, Caldirola's theory is also Lorentz invariant (continuity, in
fact,  is not an assumption required by Lorentz invariance). \ This theory
postulates the existence of a universal interval $\tau_0$ of {\em proper}
time, even if time flows continuously as in the ordinary theory.  When an
external force acts on the electron, however, the reaction of the particle
to the applied force is not continuous: The value of the electron velocity
$u_\mu$ is supposed to jump from $u_\mu(\tau - \tau_0)$ to $u_\mu(\tau)$
{\em only at certain positions} $s_{n}$ along its world line; {\em these
``discrete positions" being such that the electron takes a time $\tau_0$ for
travelling
from one position $s_{\rm{n} - 1}$ to the next one $s_{n}$.} \ The electron,
in principle, is still considered as pointlike, but the Dirac relativistic
equations for the classical radiating electron are replaced: \ (i) by a
corresponding {\em finite--difference} (retarded) equation in the velocity
$u^\mu(\tau)$

\begin{eqnarray}
{{m_0} \over {\tau_0}}\left\{ {u_\mu \left( \tau  \right)-u_\mu \left(
{\tau -\tau_0} \right)+{{u_\mu \left( \tau  \right)
u_\nu \left( \tau  \right)} \over {c^2}}\left[ {u_\nu \left( \tau
\right)-u_\nu \left( {\tau -\tau_0} \right)} \right]} \right\} \ =\nonumber \\
= \ {e \over c}F_{\mu \nu}\left( \tau  \right)u_\nu \left( \tau  \right) ,
\end{eqnarray}   

\noindent
which reduces to the Dirac equation (2) when $\tau_{0} \rightarrow 0$; \ and \
(ii) by a second equation [the {\em transmission law\/}] connecting this time
the discrete positions $x^\mu(\tau)$ along the world line of the particle:\\

\hfill{$
x_\mu \left( {n\tau_0} \right)-x_\mu \left[ {\left( {n-1} \right)\tau_0} \right]=
{{\tau_0} \over 2}\left\{ {u_\mu \left( {n\tau_0} \right)-u_\mu \left[ {\left( {n-1}
\right)\tau_0} \right]} \right\} ,
$\hfill}  (6') \\   

\noindent
which is valid inside each discrete interval $\tau_{0}$, and describes the
{\em internal} motion of the electron. \ In these equations, $u^\mu(\tau)$ is
the ordinary four-vector velocity, satisfying the condition \ $u_\mu(\tau)
u^\mu(\tau) = -c^2$ \ for \ $\tau = n \tau_0$, \ where $n = 0,1,2,...$ \ and \
$\mu,\nu = 0,1,2,3$; \ while $F^{\mu \nu}$ is the external (retarded)
electromagnetic field tensor, \ and the chronon associated with the electron
(by comparison with Dirac's equation) resulted to be\\

\hfill{$
{\tau_0 \over 2} \equiv \theta_0 = {2 \over 3}{{k e^2} \over {m_0 c^3}} \simeq
6.266 \times 10^{-24} \; {\rm s} \ ,
$\hfill} \\

\noindent
depending, therefore, on the particle (internal) properties [namely, on its
charge $e$ and rest mass $m_0$].

As a result, the electron happens to appear eventually as an
extended--like\cite{RECSAL5} particle, with internal structure, rather than
as a pointlike object.  For instance, one may imagine that the particle
does not react instantaneously to the action of an external force because
of its finite extension (the numerical value of the chronon is of the same
order as the time spent by light to travel along an electron classical
diameter). \ As already said, eq.(6) describes the motion of an object that
happens to be pointlike only at discrete positions $s_{n}$ along its
trajectory; even if both position and velocity are still continuous and
well-behaved functions of the parameter $\tau$, since they are differentiable
functions of $\tau$.  It is essential to notice that a discreteness character
is given in this way to the electron without any need of a ``model" for the
electron.  Actually it is well-known that many difficulties are met not only
by the strictly pointlike models, but also by the extended-type particle
models (``spheres", ``tops", ``gyroscopes", etc.).  We deem the answer
stays with a third type of models, the ``extended-like" ones, as the present
approach; or as the (related) theories\cite{RECSAL5} in which the
center of the {pointlike} charge is spatially distinct from the particle
center-of-mass. \ Let us repeat, anyway, that also the worst troubles met
in quantum field theory, like the presence of divergencies, are due to the
pointlike character still attributed to (spinning) particles; since
---as we already remarked--- the problem of a suitable model for elementary
particles was transported, {\em unsolved}, from classical to quantum physics.
One might say that problem to be the most important even in modern particle
physics.

Equations (6) and the following one, together, provide a full description of
the motion of the electron; but they are {\em free} from pre-accelerations,
self-accelerating solutions, and problems with the hyperbolic motion.

In the {\em non-relativistic limit} the previous (retarded) equations
get simplified, into the form

\begin{equation}
{{m_0} \over {\tau_0}}\left[ {{\bf v}\left( t \right)-{\bf v}\left(
{t-\tau_0} \right)} \right]= e \left[ {{\bf E}\left( t \right)+{1 \over c}
{\bf v}\left( t \right)\wedge {\bf B}\left( t \right)} \right] ,
\end{equation} \\  

\hfill{$
\bf r\left( t \right)-\bf r\left( {t-\tau_0} \right)={{\tau_0} \over 2}\left[ {{\bf v}\left( t \right)
-{\bf v}\left( {t-\tau_0} \right)} \right] \ ,
$\hfill} (7')  \\  

\noindent
The important point is that eqs.(6), \ or eqs.(7), \ allow to overcome the
difficulties met with the Dirac classical equation. \ In fact, the
electron {\em macroscopic} motion is completely determined once velocity and
initial position are given. \ The explicit solutions of the above
relativistic-equations for the radiating electron  ---or of the corresponding
non-relativistic equations--- verify that the following questions cab be
regarded as having been solved within Caldirola's theory: \ \
A) {\em exact relativistic solutions\/}: \ \  1) free
electron motion; \ 2) electron under the action of an electromagnetic
pulse; \ 3) hyperbolic motion; \ \ B) {\em
non-relativistic approximate solutions\/}: \ \ 1) electron under the action of
time-dependent forces; \ 2) electron in a constant, uniform magnetic
field; \ 3) electron moving along a straight line under the action
of an elastic restoring force.

In refs.\cite{RUY1} we studied the electron radiation properties as deduced
from the finite-difference relativistic equations (6), and their series
expansions, with the aim of showing the advantages of the present formalism
w.r.t. the Abraham-Lorentz-Dirac one.

\subsection{The three alternative formulations}

Two more (alternative) formulations are possible of Caldirola's equations,
based on different discretization procedures. In fact, equations
(6) and (7) describe an intrinsically radiating
particle.  And, by  expanding equation (6))
in terms of $\tau_0$, a radiation reaction term appears.  Caldirola called
those equations the {\em retarded} form of the electron equations of motion.

On the contrary, by rewriting the finite--difference equations in
the form:

\begin{eqnarray}
{{m_0} \over {\tau_0}}\left\{ {u_\mu \left( {\tau +\tau_0} \right)-u_\mu
\left( \tau \right)+{{u_\mu \left( \tau  \right)u_\nu \left( \tau  \right)}
\over {c^2}}\left[ {u_\nu \left( {\tau +\tau_0} \right)-u_\nu \left( \tau
\right)} \right]} \right\} \ =\nonumber \\
 = \ {e \over c}F_{\mu
\nu}\left( \tau  \right)u_\nu \left( \tau  \right) \ ,
\end{eqnarray} \\  

\hfill{$
x_\mu \left[ {\left( {n+1} \right)\tau_0} \right]-x_\mu \left( {n\tau_0}
\right)=\tau_0 u_\mu \left( {n\tau_0} \right) \ ,
$\hfill} (8')  \\  

\noindent
one gets the {\em advanced} formulation of
the electron theory, since the motion is now determined by advanced
actions.  At variance with the retarded formulation, the advanced one
describes an electron which absorbs energy from the external world.

Finally, by adding together retarded and advanced actions, Caldirola
wrote down the {\em symmetric} formulation of the electron theory:

\begin{eqnarray}
{{m_0} \over {2\tau_0}}\left\{ {u_\mu \left( {\tau +\tau_0} \right)-u_\mu \left(
{\tau -\tau_0} \right)+{{u_\mu \left( \tau  \right)u_\nu \left( \tau  \right)} \over {c^2}}
\left[ {u_\nu \left( {\tau +\tau_0} \right)-u_\nu \left( {\tau -\tau_0} \right)} \right]}
\right\} \ =\nonumber \\
= \ {e \over c}F_{\mu \nu}(\tau)u_\nu(\tau) \ ,
\end{eqnarray} \\  

\hfill{$
x_\mu \left[ {\left( {n+1} \right)\tau_0} \right]-x_\mu \left( {\left( {n-1} \right)\tau_0}
 \right)=2\tau_0u_\mu \left( {n\tau_0} \right) \ ,
$\hfill} (9')   \\ 

\noindent
which does not include any radiation reactions, and describes
a non radiating electron.\\

Before closing this brief introduction to the classical ``chronon theory",
let us recall at least one more result derivable from it. \ If we  consider a free
particle and look for the ``internal solutions" of the
equation (7'),  we get  ---for a periodical solution of the type

$$\dot{x}=-\beta_0 \; c \; \sin\left({\frac{2 \pi \tau}{\tau_0}}\right); \ \ \
\dot{y}=-\beta_0 \; c \; \cos\left({\frac{2 \pi \tau}{\tau_0}}\right); \ \ \
\dot{z}=0$$

\noindent
(which describes a uniform circular motion) and by imposing the kinetic energy
of the internal rotational motion to equal the intrinsic energy $m_0c^2$ of
the particle---  that the amplitude of the oscillations is given by
$\beta_0^2=\frac{3}{4}$. \ Thus, the magnetic moment corresponding to this
motion is exactly the {\em anomalous magnetic moment} of the
electron, obtained in a purely classical context: \ $\mu_a=\frac{1}{4 \pi} \;
\frac{e^3}{m_0c^2}$. This shows, by the way, that the anomalous magnetic moment
is an intrinsically classical, and not quantum, result; and the absence of
$\hbar$ in the last expression is a confirmation of this fact.

\section{Discretized Quantum Mechanics}

Let us pass to a topic we are more interested in, which is a second step
towards our eventual application of the discretization procedures for a
possible solution of the measurement problem in Quantum Mechanics, without
having to make recourse to the reduction (wave-packet instantaneous collapse)
postulate. \  Namely, let us focus our attention, now, on the consequences
for QM of the introduction of a chronon. In our (unpublished)
refs.\cite{RUY1}, we have extensively examined such consequences: Here, we
shall recall only some useful results.

There are physical limits that, even in ordinary QM, seem to prevent the
distinction of arbitrarily close successive states in the time evolution of
a quantum system. Basically, such limitations result from the Heisenberg
relations; in such a way that, if a discretization is to be introduced in the
description of a quantum system, it cannot possess a universal value
(since those limitations depend on the characteristics of the particular system
under consideration): In other words, the value of the fundamental interval
of time has to change a priori from system to system. All these points
are in favour of the extension of Caldirola's procedure to QM. \ Time will
still be a continuous variable, but the evolution of the system
along its world line will be regarded as discontinuous. In analogy with
the electron theory in the {\em non-relativistic} limit, one has to
substitute the corresponding finite--difference expression for the
time derivatives;  e.g.:

\begin{equation}
{{\drm f\left( t \right)} \over {\drm t}}\to {{f\left( t \right)-f\left( {t-\Delta t} \right)}
\over {\Delta t}} \ ,
\end{equation}  

\noindent
where proper time is now replaced by the local time $t$. \ The chronon
procedure can then be applied to obtain the finite--difference
form of the Schr\"{o}dinger equation. As for the electron case, there are
three different ways to perform the discretization, and three
``Schr\"{o}dinger equations" can be obtained:\\

\begin{equation}
i{\hbar  \over \tau }\left[ {\Psi \left( {\bf x,t} \right)-\Psi \left( {\bf x,t-\tau } \right)} \right]
= \hat{H}\Psi \left( {\bf x,t} \right) \ ,
\end{equation}  \\  

\hfill{$
i{\hbar  \over {2\tau }}\left[ {\Psi \left( {\bf x,t+\tau } \right)-\Psi \left(
{\bf x,t-\tau } \right)} \right]=\hat{H}\Psi \left( {\bf x,t} \right) \ ,
$\hfill} (11b)  \\  

\hfill{$
i{\hbar  \over \tau }\left[ {\Psi \left( {\bf x,t+\tau } \right)-\Psi \left( {\bf x,t}
\right)} \right] = \hat{H}\Psi \left( {\bf x,t} \right) \ ,
$\hfill} (11c)  \\  

\noindent
which are, respectively, the {\em retarded}, {\em symmetric} and
{\em advanced} Schr\"{o}dinger equations, all of them transforming
into the (same) continuous equation when the fundamental interval of time
(that can now be called just $\tau$) goes to zero.

Since the equations are different, the solutions they provide are also
fundamentally different. As we have already seen, in the classical theory
of the electron the symmetric equation represented a non-radiating motion,
providing only an approximate description of the motion (without taking into
account the effects due to the self fields of the electron).  However, in the
quantum theory it plays a fundamental role. In the discrete formalism too,
the symmetrical equation constitutes the only way to describe a bound
non-radiating particle. \ Let us remark that, for a time independent
Hamiltonian, the outputs obtained in the discrete formalism by using the
symmetric equation resulted to be\cite{RUY1} very similar to those obtained
in the continuous case. For these Hamiltonians, the effect of discretization
appears basically in the frequencies associated with the time dependent
term of the wave function; and, in general, seem to be negligible.

However, the solutions of the {\em retarded} (and {\em advanced\/}) equations
show a completely different behaviour. For a Hamiltonian explicitly
independent of time, the solutions have a general form given by

$$\Psi \left( {\bf x,t} \right)=\left[ {1+i{\tau  \over \hbar }\hat{H}} \right]^{{{-t}
\mathord{\left/ {\vphantom {{-t} \tau }} \right. \kern-\nulldelimiterspace} \tau }}f\left(
{\bf x} \right)$$

\noindent
and, expanding $f(x)$ in terms of the eigenfunctions of $\hat{H}$:

$$\hat H u_n\left( {\bf x} \right)=W_nu_n\left( {\bf x} \right)$$,

\noindent
that is, writing \ $f\left( {\bf x} \right)=\sum\limits_n {c_nu_n}\left(
{\bf x} \right)$, \ with \ $\sum\limits_n {\left| {c_n} \right|^2}=1$, \
one can obtain that

$$\Psi \left( {\bf x,t} \right)=\sum\limits_n {c_n}\left[ {1+i{\tau
\over \hbar }W_n} \right]^{{{-t} \mathord{\left/ {\vphantom {{-t} \tau }}
\right. \kern-\nulldelimiterspace} \tau }}u_n\left( {\bf x} \right) \ .$$

The norm of this solution is given by

$$\left| {\Psi \left( {\bf x,t} \right)} \right|^2=\sum\limits_n {\left|
{c_n} \right|}^2\exp \left( {-\gamma_nt} \right)$$

\noindent
with

$$\gamma_n={1 \over \tau }\ln \left( {1+{{\tau ^2} \over {\hbar ^2}}W_n^2}
\right)={{W_n^2} \over {\hbar ^2}}\tau +O\left( {\tau ^3} \right) \ ,$$

\noindent
where it is apparent that the damping factor depends critically on the
value $\tau$ of the chronon. \ This {\em dissipative} behaviour originates
from the character of the {\em retarded} equation; in the case of the electron,
the retarded equation possesses intrinsically dissipative solutions,
representing a radiating system. The Hamiltonian has the same status as
in the ordinary (continuous) case: It is an observable, since it is a
hermitian operator and its eigenvectors form a basis of the state
space.  However, as we have seen, the norm of the state vector is not
constant any longer, due to the damping factor. \ An opposite behaviour is
observed for the solutions of the advanced equation, in the sense that they
increase exponentially.

One of the most impressive achievement due to the introduction of
the chronon hypothesis in the realm of QM has been obtained in the
description of a bound electron by using the new formalism. In fact,
Caldirola found for the excited state of the electron the value \
$E \simeq 105.55 \;$MeV, \ which is extremely close (with an error of 0.1\%)
to the measured value of the rest mass of the {\bf muon}.  For this, and
similar questions, we just refer the reader to the quoted literature.

\section{Discretized (retarded) Liouville equation, and a solution of
the measurement problem:
Decoherence from dissipation}

Suppose we want to measure\cite{BALL6} the dynamical variable $R$ of a (microscopic)
object $\Ocal$, by utilizing a (macroscopic) measuring apparatus $\Acal$. \
The eigenvalue equation \ $R {|r\rangle}_{\Ocal} = r {|r\rangle}_{\Ocal}$ \
defines a complete eigenvector--basis for the observable $R \:$; so that
any state ${|\psi\rangle}_{\Ocal}$ of $\Ocal$ can be given by the expansion \
${|\psi\rangle}_{\Ocal} = \sum_r c_r {|r\rangle}_{\Ocal}$.

As to the apparatus $\Acal$, we are interested only in its observable
$A$, whose eigenvalues $\al$ represent the value indicated by a {\em pointer\/}; \
then, we can write \ $A {|\al,N\rangle}_{\Acal} = \al {|\al,N\rangle}_{\Acal}$, \
quantity $N$ representing the set of internal quantum numbers necessary
to specify a complete eigenvector--basis for it. \ Let the initial state of
$\Acal$ be ${|0,N\rangle}_{\Acal}$; in other words, the pointer is assumed to
indicate initially the value zero. \ The interaction between $\Ocal$ and
$\Acal$ is expressed by a time--evolution operator $U$, which is expected to
relate the value of $r$ with the measurement $\al$.\\

In conventional (``continuous") quantum mechanics, the {\em density operator},
$\rho$, obeys the Liouville--von~Neumann (LvN) equation\\

\hfill{$
\disp{{{\drm \rho} \over {\drm t}} \; = \; - {i \over \hbar} \: [H, \rho] \:
\equiv \: - i \; {\Lcal} \; \rho(t)} \ ,
$\hfill} \\

where $\Lcal$ is the Liouville operator; \ so that, if the hamiltonian $H$
is independent of time, the time evolution of $\rho$ is\\

\hfill{$
\disp{\rho (t-t_0) \; = \; \exp \left( {-{i \over \hbar}} H (t-t_0) \right) \:
\rho_0 \: \exp \left( {i \over \hbar} H (t-t_0) \right)} \ .
$\hfill}  \\

It is known that, if the compound system $\Ocal$ plus $\Acal$ is initially, for
instance,\footnote{By contrast, if we consider as initial state for the
system $\Ocal$ plus $\Acal$ the pure state \ ${|\psi^{\inr}_{N}\rangle} =
{|\psi\rangle}_{\Ocal} \bigotimes {|0,N\rangle}_{\Acal} \; \equiv \;
{|\psi\rangle}_{\Ocal} {|0,N\rangle}_{\Acal}$, \ then, within the
{\em ordinary\/} ``continuous" approach, the time evolution leads
necessarily to a coherent superposition of (macroscopically distinct)
eigenvectors: \ $U(t,t_0) \: {|\psi\rangle}_{\Ocal} {|0,N\rangle}_{\Acal} \;
= \; \sum_r c_r {|\al_r;r,N\rangle} \; \equiv \; {|\psi^{\fir}_N\rangle}$. \
As a consequence, as wellknown, one has to postulate a state collapse from
${|\psi^{\fir}_N\rangle}$ to ${|\al_{r_0}; r_0, N\rangle}$, where $r_0$ is
the value indicated by the pointer after the measurement.} in the mixed state\\

\hfill{$
\rho^{\inr} \; = \; \sum_M \, C_M \, |\psi^{\inr}_M\rangle \langle\psi^{\inr}_M| \ ,
$\hfill} \\

where quantities $C_M$ are (classical) probabilities associated with the
states ${|\psi^{\inr}_M\rangle}$, \ then the ``continuous" approach is
known to forward\\

\hfill{$
\rho^{\fir}  \equiv  U \rho U^{\dag} \; = \; \sum_M \, C_M \,
{|\psi^{\fir}_M\rangle}{\langle\psi^{\fir}_M|} \; =\\
 = \; \sum_{r_1,r_2} \, c^{*}_{r_1} c_{r_2} \, \sum_M \, C_M \, \{
{|\al_{r_1}; r_1,M\rangle} {\langle\al_{r_2}; r_2,M|} \} \ ,
$\hfill} \\

where the off-diagonal terms yield a coherent superposition of the
corresponding eigenvectors. \ In this case, the reduction postulate has to
imply that, in the measurement process, the non-diagonal terms do
instantaneously vanish.\\

{\bf On the contrary}, {\em in the discrete case}, with the interaction embedded
in the Hamiltonian $H$, {\em the situation is completely different}. \ Let us
consider the energy representation, where $|n\rangle$ are the states with
defined energy: \ $H|n\rangle=E_n |n\rangle$. \ Since the time evolution
operator is a function of the Hamiltonian, and commutes with it, the basis of
the energy eigenstates will be a basis also for this operator.

The discretized ({\em retarded\/}) Liouville--von~Neumann equation is

\begin{equation}
\disp{{ {\rho(t)-\rho(t-\tau)} \over {\tau} } \ = \ - i \, {\cal L} \; \rho(t)} \ ,
\end{equation}   

which reduces to the LvN equation when $\tau \rightarrow 0$. The essential
point is that, following e.g. a procedure similar to Bonifacio's\cite{BONI7},
one gets in this case a {\em non-unitary} time-evolution operator:

\begin{equation}
V(t,0) \; = \; {\left[ 1 + \frac{i \tau {\cal{L}}}{\hbar}
\right]}^{- t / \tau} \ ,
\end{equation}   

\noindent
which, as all non-unitary operators, does not preserve the probabilities
associated with each of the energy eigenstates (that make up the expansion
of the initial state in such a basis of eigenstates). We are interested in
the time instants $t=k\tau$, with $k$ an integer.\footnote{Let us emphasize
that the appearance of non-unitary time-evolution operators is not
associated with the {\em coarse graining} approach only, since they also
come out from the discrete Schr\"{o}dinger equations.} Thus, the
time-evolution operator (13) takes the initial density operator $\rho^{\rm in}$
to a final state for which the non-diagonal terms decay exponentially with
time; namely, to

\begin{equation}
\rho_{rs}^{\rm fin} \, = \, \langle r| V(t,0) |s\rangle \, = \,
\rho_{rs}^{\rm in} \, {\left[ {1+i\omega_{rs} \tau} \right]}^{- t/\tau} \ ,
\end{equation}    

\noindent
where

\begin{equation}
\omega_{rs} \, \equiv \, \frac{1}{\hbar} \: ({E_r - E_s}) \, \equiv \,
\frac{1}{\hbar} \: (\Delta E)_{rs} \ .
\end{equation}  

Expression (14) can be written

\begin{equation}
\rho_{rs}(t) \, = \, \rho_{rs}(0) \; \erm^{-\gamma_{rs}t} \;
\erm^{-i\nu_{rs}t} \ ,
\end{equation}  

\noindent
with

\begin{equation}
\gamma_{rs} \; \equiv \; \frac{1}{2\tau} \; \ln{\left({1+\omega_{rs}^2 \tau^2}\right)} \ ;
\end{equation}  

\begin{equation}
\nu_{rs} \; \equiv \; \frac{1}{\tau} \; \tan^{-1}{\left({\omega_{rs} \tau}\right)} \ .
\end{equation}  

\begin{figure}[!h]   
\begin{center}
  \scalebox{0.8}{\includegraphics{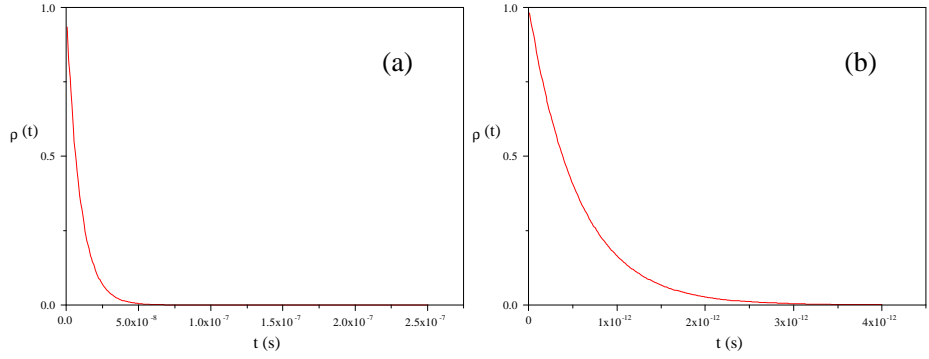}}  
\end{center}
\caption{Damping of the non-diagonal terms of the density operator for two different
values of $\tau$. For both cases we used $\Delta E=4 \;$eV. \ (a) Slower
damping for $\tau=6.26 \times 10^{-24} \;$s; (b) faster damping for
$\tau= \times 10^{-19} \;$s.}
\label{decohf1}
\end{figure}

One can observe, indeed, that the non-diagonal terms tend to zero with time,
and that the larger the value of $\tau$, the faster the decay becomes. Actually,
the chronon $\tau$ is now an interval of time related no longer to a single electron,
but to the whole system ${\cal O} + {\cal A}$. If one imagines the time
interval $\tau$ to be linked to the possibility of distinguishing
two successive, different states of the system, then $\tau$ can be
significantly larger than $10^{-23} \;$s, implying an extremely faster
damping of the non-diagonal terms of the density operator: See Fig.\ref{decohf1}.

\vfill\newpage

\newpage

\section{Comments and Conclusions}

It should be noticed that the time-evolution operator (13)  preserves trace,
obeys the semigroup law, and implies an irreversible evolution towards a
stationary diagonal form. In other words, notwithstanding the simplicity of
the present ``discrete" theory, that is, of the chronon approach, an
intrinsic relation is present between
{\em measurement process} and {\em irreversibility\/}: Indeed, the operator
(13), meeting the properties of a semigroup, does not possess in general an
inverse (and non-invertible operators are, of course, related to irreversible
processes). For instance, in a measurement process in which the microscopic
object is lost after the detection, one is just dealing with an
irreversible process that could be well described by an operator like (13).

\noindent
In our (discrete and retarded) theory, the ``state reduction"
$$ \rho (t) \vspace{1.0cm} \stackrel{t\rightarrow 0}{\rightarrow}
\vspace{1.0cm} \sum_n \rho_{nn}(0) |n\rangle \langle n| $$
\noindent
is not instantaneous, but depends ---as we have already seen--- on the
characteristic value $\tau$.  More precisely, the non diagonal terms tend
exponentially to zero according to a factor which, to the first order,
is given by
\begin{equation}
\exp{\left|{\frac{-\omega_{nm}^2 \tau t}{2}}\right|} \ .
\end{equation}  
\noindent
Thus, the reduction to the diagonal form occurs, provided that $\tau$ possesses
a finite value, no matter how small, and provided that $\omega_{nm} \tau$,
for every {\em n},{\em m}, is {\em not} much smaller than $1$; where
$$\omega_{nm} = (E_n-E_m)/\hbar$$
\noindent
are the transition frequencies between the different energy eigenstates (the
last condition being always satisfied, e.g., for non-bounded systems).

\noindent
It is essential to notice that {\em decoherence has been obtained} above,
{\em without having recourse to any statistical approach, and in particular
without assuming any ``coarse graining" of time}.  The reduction to the
diagonal form illustrated by us is a consequence of the discrete (retarded)
Liouville--von~Neumann equation only, once the inequality  $\omega_{nm} \tau
\ll 1$ is {\em not} verified.

Moreover, the measurement problem is still controversial even with regard
to its mathematical approach: In the simplified formalization introduced
above, however, we have not included any consideration beyond those common
to the quantum formalism, allowing an as clear as possible recognition
of the effects of the introduction of a chronon.

Let us repeat that the introduction of a fundamental interval of time in the
description of the measurement problem made possible a simple but effective
formalization of the state reduction process (through a mechanism that can
be regarded as a decoherence caused by interaction with the
environment\cite{DECO2})  only for the retarded case.
This is not obtainable, whem taking into account the symmetric version of the discretized
LvN equation.

It may be important to stress that the retarded form (12) of the direct
discretization of the LvN equation is {\em the same equation} obtained via
the {\em coarse grained} description (extensively adopted in \cite{BONI7}).
 \ This lead us to consider such an
equation as a basic equation for describing {\em complex systems}, which is
always the case when a measurement process is involved.

Let us add some brief remarks. {\em First\/}: the ``decoherence" does
{\bf not} occur
when we use the time evolution operators obtained directly from the retarded
Schr\"{o}dinger equation; the dissipative character of that equation, in
fact, causes the norm of the state vector to decay with time, leading again
to a non-unitary evolution operator: However, this operator
(after having defined the density matrix) yields damping terms which act also
on the diagonal terms! We discussed this point, as well a the question of
the compatibility between Schr\"{o}dinger's picture and the formalism of the
density matrix, have been analyzed by us in an Appendix of our (unpublished)
refs.\cite{RUY1}. \ {\em Second\/}: the new discrete formalism allows not only
the description of the stationary states, but also a (space-time)
description of transient states: The retarded formulation yields a
natural quantum theory for dissipative systems; and it is not
without meaning that it leads to a simple solution of the measurement
problem in QM. \ {\em Third\/}: Since the composite system ${\cal O} \; + \;
{\cal A}$ is a complex system, it is suitably described by the {\em coarse
grained} description (exploited by Bonifacio in some important papers of
his\cite{BONI7}): it would be quite useful to increase our understanding of the relationship
between the two mentioned pictures in order to get a deeper insight on the
decoherence processes involved.

\section{Acknowledgments}
One of the authors (ER) is very grateful to all the Organizers (and in
particular to Ray Chiao and Peter Milonni) of the 2002 Workshop
on Quantum Optics, for their kind invitation and for stimulating
discussions. The present paper was completed while he was visiting the ITP
of the UCSB within the activities of the mentioned Workshop: and he wishes to
thank David Gross and Daniel Hone for generous hospitality. \ The authors
are grateful, for useful discussions or kind collaboration over the years,
also to Valerio Abate, Sue Alemdar, Carlo E.Becchi, Rodolfo Bonifacio, Mauro Brambilla,
Edmundo O.Capelas, Christian Cocca, Gianni Degli Antoni, Salvatore Esposito,
Flavio Fontana, Kathy Hart, Antony Legget, Giovanni Salesi, Deborah Storm,
Michel Zamboni-Rached, Marisa T. Vasconselos, Jayme Vaz, Marco Villa and
Daniel Wisniwesky. \ A much larger presentation of the theory on which
this work is based appeared in the e-print quant-ph/9706059 [and in the
preprint IC/98/74 (ICTP; Trieste, 1998)].

\

\

\end{document}